\documentclass[conference]{IEEEtran}  

\IEEEoverridecommandlockouts                              

\usepackage{graphics} 
\usepackage{epsfig} 
\usepackage{mathptmx} 
\usepackage{times} 
\usepackage{amsmath} 
\usepackage{amssymb}  
\usepackage{subcaption}
\begin{document}
\title{\LARGE \bf BlockLoc: Secure Localization in the Internet-of-Things using Blockchain }

\author{
\IEEEauthorblockN{Omar Cheikhrouhou }
\IEEEauthorblockA{\textit{College of CIT, Taif University}\\
\textit{Taif, Saudi Arabia} \\
\textit{Computer and Embedded Systems Laboratory}\\
 \textit{University of Sfax, Sfax, Tunisia}\\
o.cheikhrouhou@tu.edu.sa}

\and

\IEEEauthorblockN{Anis Koub\^{a}a}
\IEEEauthorblockA{\textit{Prince Sultan University, Saudi Arabia} \\
\textit{CISTER/INESC-TEC, ISEP, Portugal}\\
\textit{Gaitech Robotics, China}\\
 akoubaa@psu.edu.sa}
 
 }


\maketitle
\thispagestyle{empty}
\pagestyle{empty}

\begin{abstract}
Several IoT applications are tightly dependent on the locations of the devices. However, localization algorithms can be easily compromised by injecting false locations. In this paper, we propose a Blockchain-based secure localization algorithm for the Internet of Things (IoT). The algorithm uses a public ledger (Blockchain) that contains nodes position and the list of their neighbor nodes. This ledger is shared among the IoT devices. Once an IoT device is localized its new position and the list of neighbor nodes are added to the Blockchain. This shared localization data will be used later by other IoT devices for their localization process. To avoid the attack where a malicious node sends a fake position, the correctness of the claimed position are verified before adding it to the Blockchain. Moreover, data exchanged between nodes (IoT devices) are signed to guarantee their authenticity and integrity. The integration of these security mechanisms into the localization process permits to exclude false data and therefore reduces the localization error. The simulation results show that adding the proposed security mechanism improves the localization accuracy of the algorithm when running in the presence of malicious nodes.

\end{abstract}

\section{INTRODUCTION}

According to Cisco, fifty billion devices will be deployed in the Internet-of-Things (IoT) by 2020 \cite{evans2012internet, al2018survey}. Several types of devices are connected including smartphones, healthcare sensor devices, drones and robots \cite{Dronemap, DroneTrack}, vehicles \cite{Yang2017}, industrial machines, to name a few. For all these applications, localization is essential considering that these applications are typically location-aware. In fact, localization is attracting interest due to the emerging of context-aware applications in the IoT \cite{khelifi2018survey, Koubaa:2013}. 
However, to take benefit from localization services, applications must trust the localization data and make sure that the positions are not manipulated by malicious nodes. Therefore, several research works are addressing the problem of secure localization in the Internet of Things, such as \cite{Sharma2018secLoc, chen2015securingDV, li2017securityDV,  zhang2017secure,yuan2018secure}. In fact, many localization algorithms such as those based on triangulation \cite{Koubaa:2013}, or RSSI-based algorithms \cite{JAMAA20121127, 6217729} are very dependent on the correctness of the location of nodes participating in the localization algorithm. If any node provides the wrong location information, the whole localization system will be compromised. Thus, it is crucially important to secure the localization algorithm to avoid this problem for location-aware IoT services and applications. 

\paragraph*{Related works} Secure localization in IoT is still in its infancy, and several challenges are still open. Several recent works have tackled the problem \cite{Sharma2018secLoc, chen2015securingDV, li2017securityDV,  zhang2017secure,yuan2018secure}. In \cite{Sharma2018secLoc}, the authors addressed the problem of localization of drones in urban environments, which demands high precision and accuracy in the selection of waypoints, and presented a novel solution that is capable of securing the context information for sharing 3D waypoints between UAVs. The proposed approach achieves optimal localization through hierarchical context-aware aspect-oriented Petri nets while being powered by a new drone context-exchange protocol for security validations.

The wormhole attack has been addressed in \cite{chen2015securingDV} and  \cite{li2017securityDV}.
Paper \cite{chen2015securingDV} proposed a label-based secure localization scheme to detect and defend against wormhole attack. The proposed work addressed only the wormhole attack, and so it is still vulnerable to other kinds of attacks. Moreover, the authors assumed in their network model that there is no packet loss, which is not realistic in real scenarios.

In \cite{li2017securityDV}, the authors proposed a secure localization algorithm for DV-HOP that establishes the neighbor node relationship list (NNRL) between nodes to avoid wormhole attack. All the nodes get the ID numbers of their neighbor nodes through NNRL. Then, the detection of a suspected node is done by comparing the theoretical and the actual number of neighbor nodes. These suspected nodes are eliminated from the localization process. 

The Sybil attack, where a malicious node generates several identities to hide its real identity, was addressed in \cite{yuan2018secure}. The authors proposed Sybil Free APIT (SF-APIT), a secure localization scheme for hostile distributed wireless sensor networks that can detect Sybil nodes. The detection mechanism is based on the received signal strength. To prove the correctness of their idea the APIT \cite{he2003APIT} localization algorithm was used as a reference.
In \cite{zhang2017secure}, the authors proposed the Secure Location of Things (SLOT) framework to mitigate the spoofing attack. They reformulate the location estimation problem as a stochastic censoring model and then proposed two algorithms to calculate the MLE (Maximum Likelihood Estimation) for the tag’s location. The first algorithm is based on a mixture model and the second on a time-difference-of-arrival.
The authors in \cite{perazzo2017drone} replaced all the fixed anchors with a single drone that flies through a sequence of waypoints. At each waypoint, the drone acts as an anchor and securely determines the positions. This approach completely eliminates the need for many expensive anchors. They propose three path planning algorithms that allow a drone to respectively measure and verify with a guaranteed precision a set of positions in a secure manner.

\paragraph*{Contributions} As compared to previous works on secure localization, in this paper we leverage the use of the Blockchain technology to prevent attacks that can potentially compromise the localization algorithms. In fact, Blockchain is fully decentralized, and the verification of security is performed collaboratively using trusted entities and does not rely on third parties. All these advantages lead us to design a new secure localization algorithm for IoT devices. Besides, the Blockchain technique provides security at two levels of protocol execution.  Indeed, it allows the protection of the exchanged localization data, and it guarantees the correctness of the provided localization data.



The remainder of this paper is organized as follows. Section \ref{blockchain} gives an overview of the Blockchain technology and its usage in IoT applications. Then, in Section \ref{threat}, we explain the possible threat models and attacks. Then, our proposed secure localization method is presented in Section \ref{proposed_algorithm} and its performance evaluation in Section \ref{perf}. Finally, we conclude and give some future works.


\section{The Blockchain Technology and its Usage in IoT}
\label{blockchain}
In this section, we introduce the Blockchain technology and its usage in the IoT field.
\subsection{Blockchain Background}


The Blockchain technology is a distributed \textit{ledger} shared between nodes in a peer-to-peer network. Basically, a ledger is simply a database that is maintained and updated by every node in the network. 

Each node in the network contains a copy of this ledger. The security of the Blockchain comes from the fact that blocks are cryptographically linked in a way that the alteration of one block requires the modification of all subsequent blocks in the chain. Moreover, as each node has a copy of the Blockchain, the attacker needs to make changes in at least 51\% of nodes in order to pass fake information, which makes the attack extremely much harder. 

When a node has data to send (a transaction),  it first signs it and then broadcasts it on the network. Each peer receiving this transaction first verifies the signature and then forward the transaction to other nodes in the network.
Special nodes on the network called \textit{miners} try to packet this transaction into a new block. For this purpose, first the miners verify the data, and then, they compute a valid nonce that gives a hash that satisfies a particular condition (generally that hash begins with a specific number of zeros).
The first miner that found the required nonce broadcasts this new block on the network to be added to the Blockchain.
To ensure that only valid blocks are propagated on the network, before re-transmitting the new block, a node makes extensive verification including the correctness of the nonce and the hash value and that the new block is linked to the latest block in the chain (i.e., it contains the hash value of the latest block in the chain).

\paragraph*{Consensus and Proof of Work}
In order to guarantee a specific extension rate (number of blocks added per second), and to make tampering with a block a difficult task, Blockchain system introduces a consensus protocol that defines rules for adding new blocks to the chain. The main used consensus mechanism is Proof of work (PoW). 
In the proof of work consensus protocol a miner needs to find a nonce (a random number) that produces a hash satisfying certain condition (generally that the hash is less than a threshold). The task of finding such nonce is difficult (energy and time consuming); however, its verification by other nodes is easy. The difficulty in the PoW mechanism is updated every 2016 new blocks in such a way to guarantee the desired inclusion rate (in the Bitcoin network it is fixed to 10 minutes) \cite{jesus2018survey}.

\subsection{Blockchain in IoT applications }
Blockchain is an emerging technology that is introduced in many fields especially to provide security and distributed trust between peer-to-peer nodes. In what follows, we discuss some works that leveraged the use of Blockchain technology for the security of IoT applications.
Blockchain technology was integrated into IoT to provide authentication and access control in  \cite{ouaddah2016fairaccess, jesus2018survey, hammi2018bubbles}.
In \cite{ouaddah2016fairaccess}, the authors proposed an access control mechanism based on Blockchain called \textit{FairAccess}.
The proposed solution is a fully decentralized pseudonymous and privacy-preserving authorization management framework that enables users to own and control their data. To fit their model, the authors adapted the Blockchain into a decentralized access control manager, and they used it to store the access permissions to resources.
A Blockchain-based authentication mechanism for IoT was proposed in \cite{hammi2018bubbles}. The proposed solution provides authentication of communicating things and the integrity of transmitted and stored data through the creation of secure virtual zone called \textit{bubbles of trust}. 
Before any communication can occur between two nodes of the same zone, a transaction must be transmitted and validated by this Blockchain. This rule presents a main weakness of this solution as it introduces a big latency (depends on the inclusion rate the Blockchain) in the communicating system. 


The authors in \cite{kchaou2018toward}  proposed a distributed trust management scheme for VANET security based on clustering and Blockchain. Before adding a new block, the scheme requires the verification of the correctness of the message based on the vehicle behavior which is controlled by the miner and the credibility of the message decided by a Cluster Header.

In \cite{dorri2017blockchain}, Dorri et al. addressed the heavy computation load of the Blockchain technology and provided a lightweight Blockchain solution for IoT to secure smart home. The proposed solution eliminates the  Proof of work and cryptocurrencies concepts.

\section{Threat Model for Secure Localization}
\label{threat}
Malicious and fake nodes in the IoT positioning system could intentionally send corrupted or fake information to disturb the localization system. Several attacks with various impact target the localization scheme in IoT, including:

\begin{itemize}
  \item \textit{\textbf{Eavesdropping the devices position}:}  in some localization systems the position of devices is sent to other nodes. If this information is sent without encryption, outsider attacker might eavesdrop the communication and disclose the IoT device position. This fact breaches the privacy of the user and makes the confidentiality of the user in danger \cite{cheikhrouhou2016secure, cheikhrouhou2012lnt}.
  \item \textit{\textbf{Message forging}:} when a node sends its position in the network, an attacker can intercept the message and forge it by putting a false position. This behavior could disturb the whole localization system in the network \cite{Cheikhrouhou:2011:RRB}.
  \item \textit{\textbf{Wormhole attack}:} In the wormhole attack, two malicious nodes (called \textit{wormhole nodes}), strategically placed at distant regions, collude to create a wormhole link. This wormhole link can be created using out-of-band or even wired link. Through this wormhole link the malicious nodes make victims  (called \textit{affected nodes}), in one region, believe that they are close to the far apart nodes in the distant region, which is a deceptive belief to attract and sniff victims data \cite{xiao2010handbook}. The wormhole attack has a significant impact especially for localization mechanisms based on RSSI or on the topology the network \cite{chen2015securingDV}. Paper \cite{chen2015securingDV} explains the impact of the wormhole attack at the DV-Hop localization algorithm.
  \item \textit{\textbf{Sybil attack}:} In this attack, the malicious node illegitimately presents several addresses to hide its real identity or to gain more access to the network resources \cite{patel2017reviewSybil}. The Sybil attack has several forms; fabricated identities, stolen identities, simultaneous and non-simultaneous, etc. The Sybil attack has a severe impact on the localization of nodes and might totally disturb the operation of the network. The impact of this attack on the localization system was explained in \cite{yuan2018secure}.

\end{itemize}

\begin{figure*}[htb]
    \centering
    \includegraphics[width=0.5\textwidth]{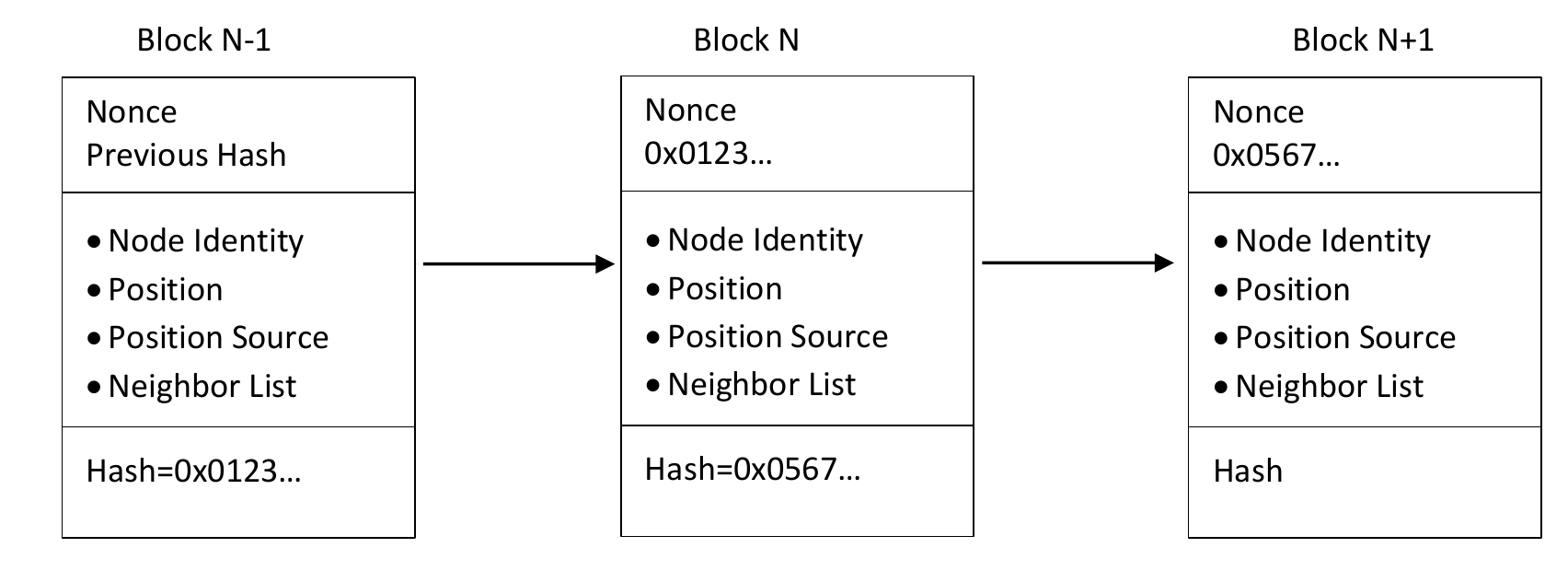} %
    \caption{The BlockLoc BlockChain Structure}
    \label{fig:blockStruct}
\end{figure*}

\section{BlockLoc: Blockchain Localization Algorithm for IoT Applications}
\label{proposed_algorithm}
\subsection{System Model and Assumptions}
In our application, we consider a set of IoT devices (called nodes) that collaborate together to determine their position. In our model, we have the following requirements:
 \begin{itemize}
   \item \textbf{Decentralization}: In our model, there is no central entity that computes the localization position of nodes: All nodes are peers that collaborate in the localization system.
   \item \textbf{P2P Communications}: In our proposed localization scheme, the IoT devices communicate with its neighbors' nodes to determine their positions, in a peer-to-peer network architecture.
   \item \textbf{No Central Trust}: The proposed network model does not require a central trusted entity that manages the security between nodes or detects the existing of malicious nodes. The trust is provided thanks to the use of the Blockchain technology.
 \end{itemize}

The aforementioned characteristics of an IoT network model make the use of a Blockchain a necessity as this latter provides a secure distributed ledger and can ensure a distributed trust between the different peers IoT devices.
Moreover, a successful security protocol needs to fit the constrained resources of IoT devices. Indeed, IoT networks generally consist of heterogeneous devices such as smartphone, watch, wireless sensor nodes, etc. These later have low computation, low memory, and low energy power \cite{cheikhrouhou2010lightweight}. Therefore, security protocols need to be lightweight and low-power consuming. 
\subsection{The BlockLoc Algorithm}

Blockchain is considered as a framework to secure the data exchanged between a set of peer nodes and to provide trust between them. In our work, we use Blockchain for two purposes: (1) first to protect the localization data exchanged between IoT devices, (2) second to guarantee the correctness of the given position data.


In what follows, we propose BlockLoc, a secure localization scheme that uses the Blockchain technology to protect the exchanged localization information and to guarantee the correctness of the claimed node's position.
In the BlockLoc localization method, nodes collaborate to determine their position. More precisely, a node needs to communicate with at least three anchors (i.e., nodes with known positions) in order to determine its position. Triangulation is an example of a localization technique that uses at least three anchors to determine the location of a fourth node. BlockLoc is algorithm agnostic, which means it can be applied to any distributed localization algorithms, that is based on location data exchange. 
However, instead of sending positions in an unprotected message that could be forged by attackers, the node gets the neighbors positions from the secure Blockchain ledger. This permits avoiding forging attacks. Moreover, a malicious node can provide a fake position to disturb the network. To mitigate against this behavior, every claimed node position is verified before adding it to the Blockchain ledger. For this purpose, the claimed node is required to send, in addition to its position, the list of its neighbor nodes. By verifying that the list of neighbor nodes are really in the vicinity of the claimed node, the localization scheme can exclude malicious data.

We assume that each node has two keys: (\textit{i.}) one key is public and known by all nodes and (\textit{ii.}) the other key is kept private and secret. 

IoT applications generally use heterogeneous devices. Some devices can be equipped with GPS and so can determine their positions (e.g. smartphone, car, etc.). Other devices might not be equipped with a GPS, and so they need to run the proposed secure localization scheme to determine their positions. In order to work properly, the BlockLoc scheme requires that a node knows the positions of at least three nodes and the corresponding distance to them. These nodes will play the role of anchors. These anchor nodes can be either neighbors or not. If the anchor node is neighbor, the RSSI technique is used to estimate the distance between the node and this anchor \cite{Koubaa:2013}. Otherwise, the DV-Hop technique is used \cite{cheikhrouhou2018hybrid}. Note that RSSI stands for \textit{Received Signal Strength Indicator} and is used as a link quality estimator in wireless communication \cite{Baccour:2012} and also used to estimate the distance between two nodes. 

The distance between two nodes can be deduced from the received signal power of nodes using the following equation \cite{xu2017efficient}:
\begin{equation}\label{eq:rssi}
RSS(d)(dBm) = P_{tr} - P_{loss}(d_{0}) - 10\tau \log_{10} \frac{d}{d_0}+ X_{\sigma},
\end{equation}
where $d$ means the distance between the transmitting  and receiving nodes, $RSS(d)$ indicates the  signal power as received at a node located across a distance of $d$ from the transmitting node, $d_{0}$ is the reference distance, $P_{tr}$ denotes the transmitted signal's power, $P_{loss}(d_{0})$ means the signal power loss across the reference distance $d_{0}$, $\tau$ is the path loss exponent whose value depends on the medium of propagation, and $X_{\sigma}$ is the noise, which is described as a Gaussian random variable with zero as its mean and $\sigma$ as the standard deviation.
For more information about the RSSI and the DV-Hop methods, the reader can refer to the work \cite{cheikhrouhou2018hybrid}. 

More precisely, the proposed BlockLoc secure localization scheme works as follows. 
\subsubsection{Initialization} 
In our network model, we suppose the existence of fixed anchors with known positions. These anchors are generally relatively powerful nodes that could play the role of miners; nodes responsible for adding new blocks to the Blockchain.
Therefore, at the initialization phase, new blocks containing the anchors' positions are added to the Blockchain. 
These initial anchors' positions serve in the verification of the firstly localized nodes' positions.

\subsubsection{Blockchain construction} 
First, each node knowing its position adds it to the Blockchain. For this purpose, it creates a block containing; its address, position and the list of neighbor nodes, then it broadcasts this block to the network.  
Figure \ref{fig:blockStruct}, shows the Blockchain structure.

The message is sent signed. This means that the node computes a digital signature of the message using its private key. Other nodes in the network verify the security of the message by checking this signature using the sender public key. This guarantee the authenticity (the message is actually sent by the claimed node) and the integrity (the data has not been altered during transmission) of the message. Moreover, it avoids the Sybil attacks and the identity usurpation attacks as each public key is associated with one address (a node cannot claim to have different identity). More precisely, the identity of a node is the hash value of its public key.

When miner nodes receive the new block, they first verify the message signature using the sender public key. Then, the miners verify the correctness of the claimed position. This latter verification consists in verifying that the claimed position is in the vicinity of the given neighbor nodes. More precisely, they verify that the distance between the claimed position and the position of a neighbor node is less or equal to the value of the communication range.
If one of these verification operations fails, the block is ignored and the node position is excluded from the localization system.
In case both verification operations succeed, miner nodes compete to find the good nonce (the nonce that satisfies the required PoW consensus). The first miner that computes the required nonce broadcast the new block to the network. The new block contains in addition to the received data (node address, position and neighbor list) a hash code and a copy of the hash code of the last block in the chain (this permit to link blocks between them).

\subsubsection{Node Localization} 
In most existing localization schemes, the localization is based on the existence of anchor nodes. More precisely,  a node with unknown location (we call it\textit{ unknown node}) needs to have a connection to at least three anchor nodes in order to be localized. The lack of enough anchors nodes leads to the failure of the unknown node localization.
To avoid this limit, In the proposed BlockLoc secure localization scheme, an unknown node can serve by the already localized nodes. 

This permits the localization of the node even without the availability of fixed anchor nodes.
Although it can exist several localized nodes, the unknown node prioritizes the closest ones.
For this purpose, the unknown node first collects \textit{1-hop neighbor} positions ({1-hop neighbor} are the nodes that are one hop far from the unknown node). If it does not get at least three neighbor's responses, it makes a new round and contact 2-hop neighbor nodes and so on, the number of neighbor hops will be incremented at each new round until receiving at least three responses.
More precisely, when a node A wishes to be localized, first it sends a \textit{discover-message}  to its neighbors with \textit{hopcount} value equal to 1.  The \textit{hopcount} field permits to decide the number of hops the message is traveling in the network. Each neighbor node receiving this message and that is already localized, responds by sending a message containing its \textit{identity}. 
By receiving this response message the node A, first extracts the corresponding neighbor position from the Blockchain. Then,  the node A computes the distance to this neighbor node using the RSSI method. The RSSI method is used as the neighbor is a one-hop neighbor.
If the node A receives at least three responses from three neighbors, it estimates its positions using the Triangulation method \cite{hartley1997triangulation}.
Otherwise, the node A starts a new round and sends a new \textit{discover-message}  with \textit{hopcount} value equal to 2. This process of incrementing the \textit{hopcount} value is repeated until the node A receives at least three responses from already localized nodes. 
When the responder node is a one-hop neighbor to the node A the RSSI method is used; However, when the node is not directly connected to the node A, the RSSI method cannot be used and so the DV-Hop method is used.
The idea of the DV-Hop message is to first compute the average hop-distance, then the estimated distance will be equal to the average hop distance multiplied by the number of hops \cite{cheikhrouhou2018hybrid}.

\begin{figure*}[htbp]
    \centering
    \begin{subfigure}[t]{0.35\textwidth}
        \includegraphics[width=\textwidth]{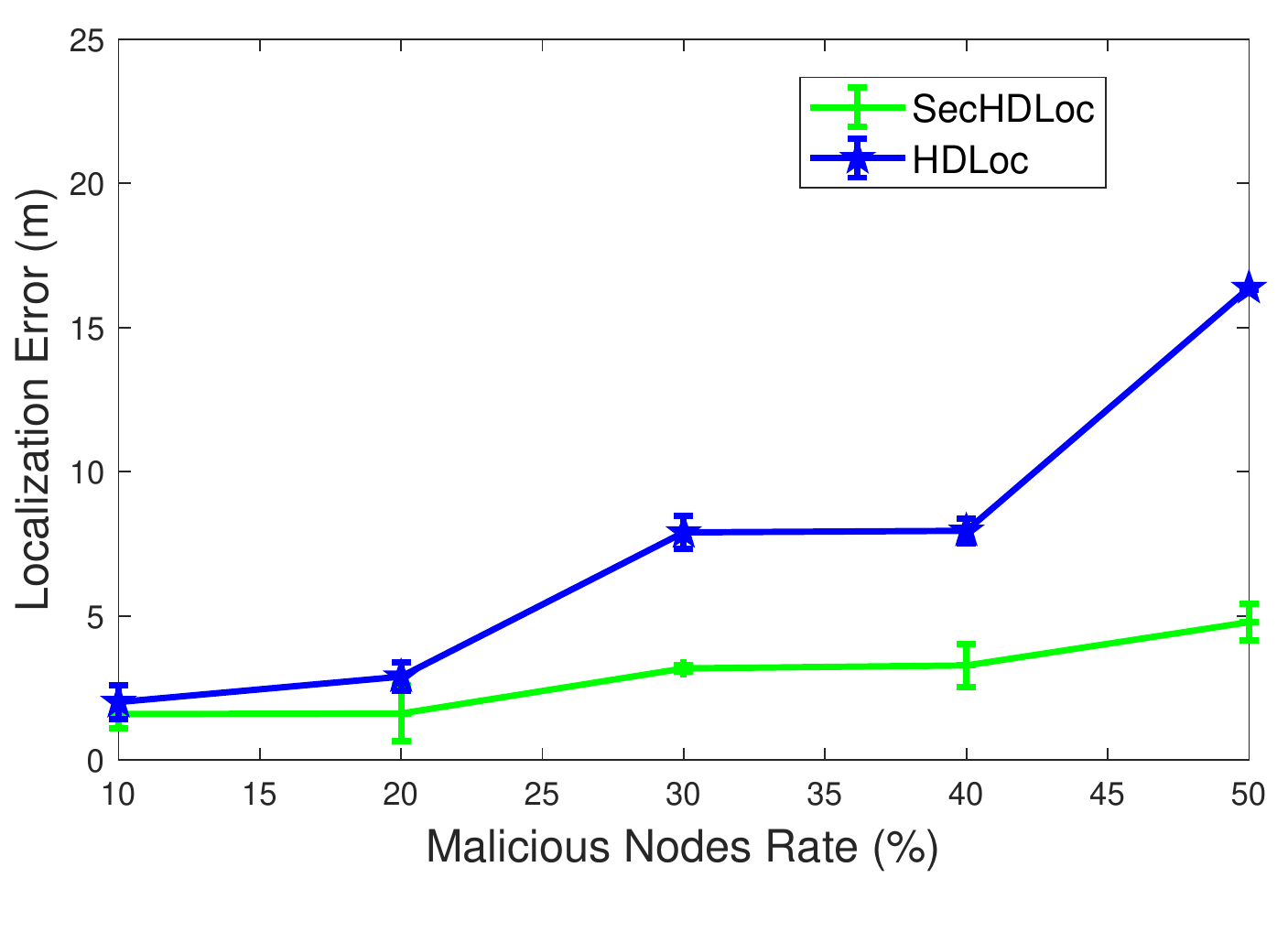} %
        \caption{Anchor rate 20\%}
        \label{subfig:LocaErrMal20}
    \end{subfigure}
    \begin{subfigure}[t]{0.35\textwidth}
        \includegraphics[width=\textwidth]{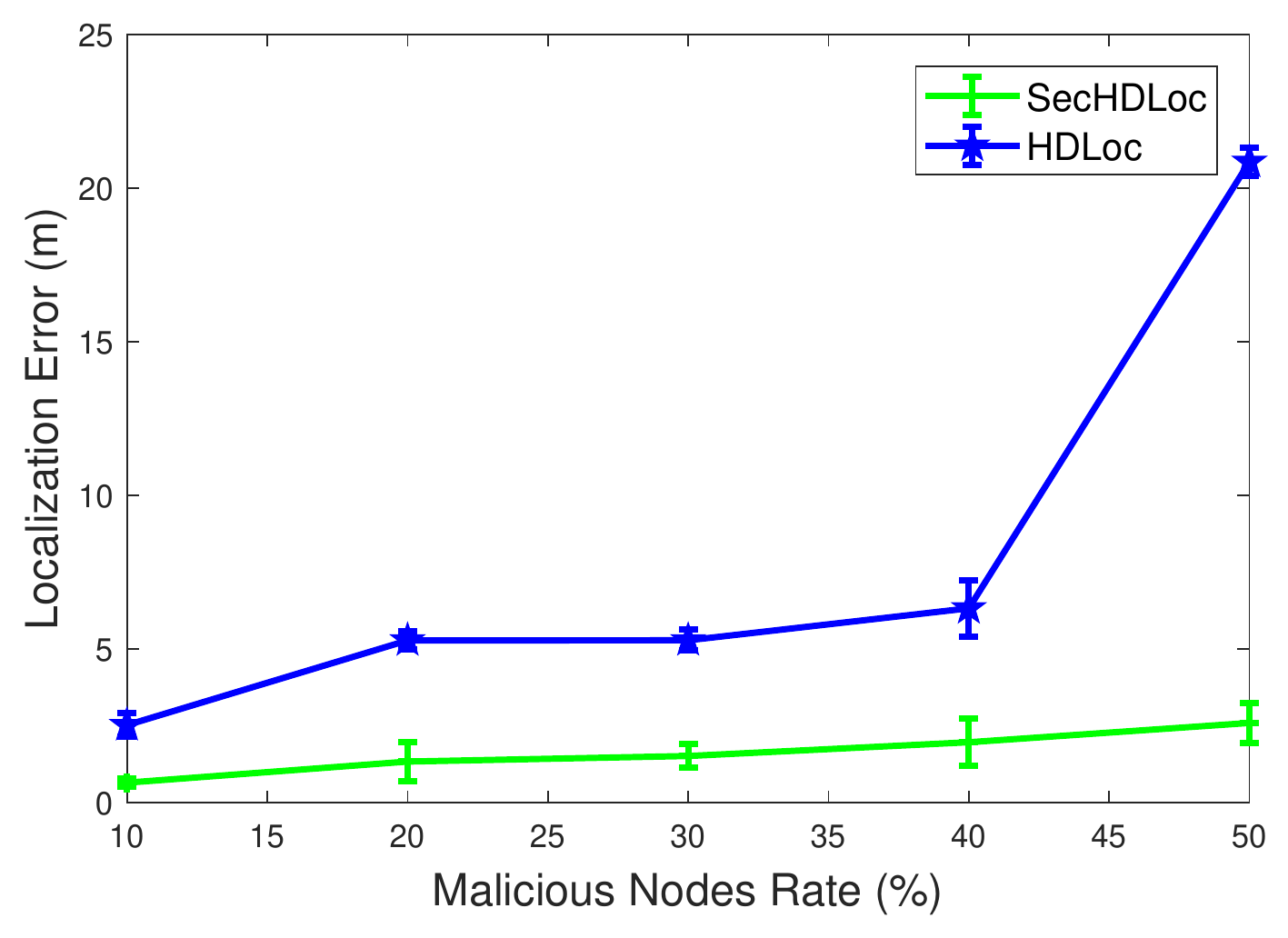} %
        \caption{Anchor rate 50\%}
        \label{subfig:LocaErrMal50}
    \end{subfigure}
    \caption{Localization error vs Malicious nodes rate}
    \label{fig:LocaErrMal}
\end{figure*}

As it can be noticed, in our proposed localization method, the nodes' positions are not sent in the network. Instead, the identities of nodes are sent along with their positions are extracted from the Blockchain. This characteristic permits to avoid the eavesdropping attack and preserve the privacy of the users.

\section{Performance Evaluation}
\label{perf}
This section presents the performance of the proposed secure localization scheme and the impact of the security improvement on the accuracy of the localization scheme under the presence of malicious nodes. In the BlockLoc proposed method, the security mechanism is based on the use of the Blockchain. To highlight the impact of this security mechanism, we have implemented two versions of the localization method. One version without any security scheme (called HDLoc for Hybrid DV-Hop Localization, which is a previously published improvement of the DV-Hop algorithm) and the second version with BlockLoc security mechanism (called SecHDLoc for Secure Hybrid DV-Hop Localization). Furthermore, we have considered malicious nodes, and we tested the impact of these nodes on the performance of the localization scheme. The malicious behavior that we take into consideration is the modification of nodes' positions. More precisely, the malicious node sends a modified value of its position. In our simulation, we introduce an error value of 50\% which means that: 
\begin{equation}
Malicious \: position=1.5 \: x \: real \: position,
\end{equation}

\subsection{Simulation Model}
In our simulation, we have used a wireless sensor network consisting of a fixed number of sensor nodes being \(100\). These nodes were randomly deployed in an area of \(100 x 100 \: m^2\). We assume that all the nodes in the network have the same characteristics. We also assume symmetric links among neighboring nodes, i.e., if node $A_i$ can receive a packet transmitted by $A_j$, then vice versa is also true. We used Matlab for implementing our simulations.
The communication range between nodes is fixed to \(30 \: m\).
During these simulations, the number of malicious nodes vary in (10\%, 20\%, 30\%, 40\%, 50\%).  For each simulation scenario, we repeat the experiment ten times with new randomly generated nodes locations.

\subsection{Simulation Results}
Figure \ref{fig:LocaErrMal} shows the impact of the increase of the number of malicious nodes on the localization accuracy.
More precisely, we have considered two scenarios. In the first scenario, the anchor rate is equal to 20\%, and in the second scenario, the anchor rate is equal to 50\%. In a real scenario, the anchors are simply IoT devices with a fixed GPS location.

In Figure \ref{subfig:LocaErrMal20}, where the number of anchor rate is 20\%, we notice that the secure version of the localization algorithm (SecHDLoc) is slightly affected by the number of malicious nodes. This is due to the fact that malicious nodes are detected and eliminated from the localization process. However, as it can be seen in Figure \ref{subfig:LocaErrMal20}, the basic version of the algorithm (HDLoc) is sharply is affected by the number of malicious node and the localization error reached almost \(16 \: m\) when the number of malicious nodes is 50\%, whereas it is only \(4 \: m\) in the secure version. \\
The increase of the anchor rate, Figure \ref{subfig:LocaErrMal50}, improves the accuracy of the secure version of the algorithm, however, it decreases the accuracy of the insecure version of the algorithm. This can be explained by the fact that the increase of anchor rate also increase malicious anchor rate (as malicious node are taken randomly and can be an anchor rate), and the error introduced by an anchor node has more impact than the error introduced by other nodes.

\section{Conclusion}
In this paper, we have proposed a secure localization scheme based on Blockchain. The proposed algorithm takes advantage of the distributed and the decentralized characteristic of Blockchain to provide a trustful framework of information sharing between nodes. Thanks to Blockchain, a malicious node cannot inject fake data to the localization mechanism as all data need to be verified and checked before adding it to the Blockchain.
The performance evaluation demonstrates the improvements of the security mechanisms on the accuracy of the localization algorithm under the presence of malicious nodes. More precisely, the introduced security mechanisms minimize the localization error to the \(1/4\) as compared to the non-secure version under the presence of 50\% of malicious nodes.
As future work, we will implement the proposed secure localization scheme on a real platform using Raspberry Pi devices.




\section*{ACKNOWLEDGMENT}
This work is supported by the Robotics and Internet of Things (RIOTU) Lab of Prince Sultan University.

                        
\bibliography{iotRef}
\bibliographystyle{IEEEtran}

\end{document}